\documentclass[12pt]{article}
\def \b{{\cal B}}
\def \beq{\begin{equation}}
\def \eeq{\end{equation}}

\begin{document}
\begin{flushright}
SLAC-PUB-12042\\
EFI 06-16 \\
hep-ph/0608040 \\
August 2006 \\
\end{flushright}
\centerline{\bf RATE AND CP-ASYMMETRY SUM RULES IN $B \to K \pi$
\footnote{Research supported in part by the US Department of Energy,
contract DE-AC02-76SF00515}}
\bigskip
\centerline{Michael Gronau$^{a,b}$ and Jonathan L. Rosner$^c$}
\medskip
\vskip3mm
\centerline{$^a$\it Physics Department, Technion -- Israel Institute of
Technology}
\centerline{\it 32000 Haifa, Israel}
\medskip
\centerline{$^b$\it Stanford Linear Accelerator Center, Stanford University}
\centerline{\it Stanford, CA 94309, USA}
\medskip
\centerline{$^c$\it Enrico Fermi Institute and Department of Physics,
University of Chicago}
\centerline{\it Chicago, Illinois 60637, USA}
\bigskip
\centerline{ABSTRACT}
\medskip
The observed violation of $A_{CP}(B^0 \to K^+ \pi^-) = A_{CP}(B^+ \to K^+
\pi^0)$ has been recently mentioned as a puzzle for the standard model. We
point out that while this violation may be accounted for by a large
color-suppressed tree amplitude, a sum rule involving three or four $B\to K\pi$
CP asymmetries should hold.  The current experimental status of these sum rules
and of a sum rule for $B\to K\pi$ decay rates is presented.

\bigskip
Recently \cite{DiMarco,Unno} the fact that $A_{CP}(B^0 \to K^+
\pi^-) \ne A_{CP}(B^+ \to K^+ \pi^0)$ was mentioned as a puzzle for
the Standard Model.  The equality of these two CP asymmetries was
proposed eight years ago \cite{Gronau:1998ep} in the limit that only
penguin ($P$) and color-favored tree ($T$) amplitudes contributed to
these decays.  Since then it has been recognized for some time
(e.g., through detailed flavor-SU(3) fits of $B$ decays to two
charmless pseudoscalar mesons \cite{Chiang:2004nm}) that the
color-suppressed ($C$) tree amplitude also plays an important role
in $B^+ \to K^+ \pi^0$ decays.  When this amplitude is included in
the discussion, a more exact sum rule was proposed
\cite{Gronau:2005gz}:
\beq \label{eqn:one} A_{CP}(K^+ \pi^-) =
A_{CP}(K^+ \pi^0) + A_{CP}(K^0 \pi^0)~~,
\eeq
or, taking account of
a small annihilation amplitude ($A$) as well \cite{Gronau:2005kz},
\beq
\label{eqn:two} A_{CP}(K^+ \pi^-) + A_{CP}(K^0 \pi^+) =
A_{CP}(K^+ \pi^0) + A_{CP}(K^0 \pi^0)~~.
\eeq
These relations also hold approximately in the presence of an electroweak
contribution $P_{EW}$, and the second can be derived using isospin
\cite{Gronau:2005kz,Atwood:1997iw}. Rather than expressing a
discrepancy with the Standard Model, they serve as an important {\it
test} of it once the CP asymmetry in $B^0 \to K^0 \pi^0$ is measured
with sufficient accuracy \cite{Hara,Hazumi}.

Eqs.~(\ref{eqn:one}) and (\ref{eqn:two}) are derived in the limit of the
leading-order ($P$) contributions to decay rates.  More accurate versions are
expressed in terms of {\it rate differences}
\beq
\Delta_{ij} \equiv \Gamma(B \to K^i \pi^j) - \Gamma(\bar B \to K^{\bar i}
\pi^{\bar j})~~~.
\eeq
Neglecting the annihilation amplitude $A$ one finds \cite{Gronau:2005gz}
\beq \label{eqn:three}
\Delta_{+-} \simeq 2(\Delta_{+0} + \Delta_{00})
\eeq
while including $A$ one has \cite{Gronau:2005kz}
\beq \label{eqn:four}
\Delta_{+-} + \Delta_{0+} \simeq 2(\Delta_{+0} + \Delta_{00})
\eeq
At the moment the CP asymmetry $A_{CP}(K^0 \pi^0)$ agrees well with the
nearly-identical predictions of Eqs.\ (\ref{eqn:three}) and (\ref{eqn:four}).
A corresponding sum rule relating the {\it rates} for the four $B \to K \pi$
processes is now seen to be satisfied at the $1 \sigma$ level.

We use the latest measured branching ratios and asymmetries from BaBar
\cite{DiMarco,Bona} and Belle \cite{Unno} summarized in Tables I and II,
respectively.  CP asymmetries, by convention, are defined in terms of
the rate differences $\Delta_{ij}$ by
\beq
A_{CP}(K^i \pi^j) \equiv -\Delta_{ij}/[\Gamma(B \to K^i \pi^j) +
\Gamma(\bar B \to K^{\bar i} \pi^{\bar j})]~~~.
\eeq

We first write the sum rule (\ref{eqn:four}) for rate asymmetries, which, when
expressed in terms of CP asymmetries, reads \cite{Gronau:2005kz}
$$
A_{CP}(K^+ \pi^-) + A_{CP}(K^0 \pi^+) \frac{\b(K^0 \pi^+)}{\b(K^+ \pi^-)}
\frac{\tau_0}{\tau_+} =
$$
\beq
A_{CP}(K^+ \pi^0) \frac{2 \b(K^+ \pi^0)}{\b(K^+ \pi^-)} \frac{\tau_0}{\tau_+}
+ A_{CP}(K^0 \pi^0) \frac{2\b(K^0 \pi^0)}{\b(K^+ \pi^-)}~~~.
\eeq

\renewcommand{\arraystretch}{1.4}
\begin{table}
\caption{Branching ratios for $B \to K \pi$ presented at ICHEP06 and their
averages, in units of $10^{-6}$.
\label{tab:br}}
\begin{center}
\begin{tabular}{c c c c} \hline \hline
Mode & BaBar \cite{Bona} & Belle \cite{Unno} & Average \\ \hline
$K^+ \pi^-$ & $19.7 \pm 0.6 \pm 0.6$ & $20.0 \pm 0.4^{+0.9}_{-0.8}$ &
 $19.83 \pm 0.63$ \\
$K^+ \pi^0$ & $13.3 \pm 0.56 \pm 0.64$ & $12.4 \pm 0.5^{+0.7}_{-0.6}$ &
 $12.83 \pm 0.59$ \\
$K^0 \pi^+$ & $23.9 \pm 1.1 \pm 1.0$ & $22.9^{+0.8}_{-0.7} \pm 1.3$ &
 $23.40 \pm 1.06$ \\
$K^0 \pi^0$ & $10.5 \pm 0.7 \pm 0.5$ & $9.2^{+0.7+0.6}_{-0.6-0.7}$ &
 $9.89 \pm 0.63$ \\ \hline \hline
\end{tabular}
\end{center}
\end{table}

\begin{table}
\caption{CP asymmetries for $B \to K \pi$ presented at ICHEP06 and their
averages.
\label{tab:as}}
\begin{center}
\begin{tabular}{c c c c} \hline \hline
Mode & BaBar \cite{DiMarco,Aubert:2006gm} & Belle \cite{Unno} & Average \\
\hline
$K^+ \pi^-$ & $-0.108 \pm 0.024 \pm 0.007$ & $-0.093 \pm 0.018 \pm 0.008$ &
 $-0.099 \pm 0.016$ \\
$K^+ \pi^0$ & $0.016 \pm 0.041 \pm 0.010$ & $0.07 \pm 0.03 \pm 0.01$ &
 $0.050 \pm 0.025$ \\
$K^0 \pi^+$ & $-0.029 \pm 0.039 \pm 0.010$ & $0.03 \pm 0.03 \pm 0.01$ &
 $0.007 \pm 0.025$ \\
$K^0 \pi^0$ & $-0.20 \pm 0.16 \pm 0.03$ & $-0.05 \pm 0.14 \pm 0.05$ &
 $-0.12 \pm 0.11$ \\ \hline \hline
\end{tabular}
\end{center}
\end{table}

Here we have converted ratios of branching ratios to ratios of rates where
necessary using the ratio $\tau_+/\tau_0 = 1.076 \pm 0.008$ of $B^+$ and
$B^0$ lifetimes \cite{HFAG}.  The sum rule (\ref{eqn:three}) is evaluated by
omitting the term containing the very small CP asymmetry $A_{CP}(K^0 \pi^+)$.
Using the averaged branching ratios and CP asymmetries in Tables I and II, we
predict
\beq
{\rm Eq.}~(\ref{eqn:four}) \Rightarrow A_{CP}(K^0 \pi^0) = -0.151 \pm 0.043~,
\eeq
\beq
{\rm Eq.}~(\ref{eqn:three}) \Rightarrow A_{CP}(K^0 \pi^0) = -0.159 \pm 0.036~,
\eeq
to be compared with the observed value
\beq
A_{CP}(K^0 \pi^0) = -0.12 \pm 0.11~~.
\eeq
Either prediction is consistent with the observed value.  Nearly identical
predictions of $A_{CP}(K^0 \pi^0) = (-0.142 \pm 0.039, -0.149 \pm 0.030)$ are
obtained using Eqs.~(\ref{eqn:two}) and (\ref{eqn:one}), respectively.

The rate sum rule \cite{Gronau:1998ep,HJL}
\beq
\Gamma(K^+ \pi^-) + \Gamma(K^0 \pi^+) = 2[\Gamma(K^+ \pi^0) + \Gamma(K^0
\pi^0)]~~,
\eeq
where isospin-breaking corrections are suppressed by a ratio of tree and
penguin amplitudes~\cite{Gronau:2006eb}, may be expressed in terms of branching
ratios by correcting for the lifetime ratio:
\beq
\b(K^+ \pi^-) + \b(K^0 \pi^+) \frac{\tau_0}{\tau_+} =
 2[\b(K^+ \pi^0) \frac{\tau_0}{\tau_+} + \b(K^0 \pi^0)]~~~.
\eeq
In units of $10^{-6}$, the left-hand side is $41.58 \pm 1.18$, while the
right-hand side is $43.63 \pm 1.68$.  The difference is $2.05 \pm 2.05$, or $1
\sigma$.  Both this sum rule and the rate difference sum rule (5) are useful
tests for new physics in the $b \to s$ penguin diagram, which has shown
hints of exhibiting new contributions elsewhere \cite{Hazumi,MGCapri}.

Note added:  The rate sum rules are dominated by a common penguin contribution
for which they are trivially satisfied.  They may be rearranged so that each
side is an interference term between the dominant penguin and subdominant
color-favored or color-suppressed tree contributions \cite{Lipkin:2005sn}.
They are, of course, still satisfied in this form, but present experimental
errors are still too large to tell whether each side of the sum rule is
nonzero with sufficient significance.

\medskip
We thank H. J. Lipkin for helpful comments.
M.G. is grateful to the SLAC Theory Group for its kind hospitality.
This work was supported in part by the Israel Science Foundation
under Grant No.\ 1052/04, by the German-Israeli Foundation under
Grant No.\ I-781-55.14/2003, and by the U. S. Department of Energy under
Grant No.\ DE-FG02-90ER40560.

\end{document}